# Measurement of high exciton binding energy in the monolayer transition-metal dichalcogenides $WS_2$ and $WSe_2$

A. T. Hanbicki[a], M. Currie[a], G. Kioseoglou[b], A. L. Friedman[a], and B. T. Jonker[a*]

[a]*Naval Research Laboratory*, 4555 Overlook Ave. SW, Washington, DC 20375
[b]*Dept. of Materials Science and Tech., University of Crete*, Heraklion Crete, 71003, Greece

Monolayer transition-metal dichalcogenides are direct gap semiconductors with great promise for optoelectronic devices. Although spatial correlation of electrons and holes plays a key role, there is little experimental information on such fundamental properties as exciton binding energies and band gaps. We report here an experimental determination of exciton excited states and binding energies for monolayer $WS_2$ and $WSe_2$. We observe peaks in the optical reflectivity/absorption spectra corresponding to the ground- and excited-state excitons (*1s* and *2s* states). From these features, we determine lower bounds free of any model assumptions for the exciton binding energies as $E_{2s}^A - E_{1s}^A$ of 0.83 eV and 0.79 eV for $WS_2$ and $WSe_2$, respectively, and for the corresponding band gaps $E_g \geq E_{2s}^A$ of 2.90 and 2.53 eV at 4K. Because the binding energies are large, the true band gap is substantially higher than the dominant spectral feature commonly observed with photoluminescence. This information is critical for emerging applications, and provides new insight into these novel monolayer semiconductors.

* Corresponding Author. Tel. (202) 404-8015; fax: (202) 404-4637

*E-mail address:* Jonker@nrl.navy.mil

1. Introduction

Transition-metal dichalcogenides (TMDs) are layered compounds like graphite that can be readily reduced to two-dimensional (2D) form due to their strong in-plane bonding and weak interlayer van der Waals coupling [1, 2, 3]. They are of keen interest for emerging electronic applications [4,5,6,7,8]. In contrast with their indirect gap character in the bulk, single monolayer TMDs of the family $MX_2$ ( M = Mo, W and X = S, Se, Te) are direct-gap semiconductors [9,10,11] suitable for a host of optoelectronic [2], light harvesting [12,13], chemical sensor [14,15], and spin-valley coupling based [16,17,18] applications. Despite recent progress, there are many gaps in our understanding of the fundamental interactions which lead to the striking optical and electronic properties these monolayer semiconductor materials exhibit, including very basic electron-hole interactions.

Photoexcitation in semiconductors is usually performed by promoting an electron directly from the valence band to the conduction band. A subsequent coulomb interaction, mitigated by dielectric screening from the surrounding lattice, leads to spatial confinement of the electron-hole pair. The resulting bound quasiparticle is known as an exciton. In bulk materials, excitons have binding energies less than 0.1 eV (*e.g.* ~ 40 meV in bulk $MoS_2$), so that the ground state of the exciton is just below the true band gap energy determined by the band structure. In reduced dimensions as in single layers of $MX_2$, dielectric screening is greatly reduced, which is predicted to lead to exciton binding energies ten times larger than in the bulk [19,20,21,22]. These high binding energy excitons exhibit behavior similar to excited atoms, with energy levels designated *1s, 2s*, *2p* and so on.

Here we report observation of *2s* excited exciton levels in monolayer $WS_2$ and $WSe_2$. The observation of both ground and excited states allows us to determine a lower bound for the exciton binding energy. We measure the optical reflectivity/absorption from several single monolayer samples in the energy range of 1.5 – 3.3 eV and temperatures of 4 – 300 K. For each sample, we observe peaks corresponding to the ground state exciton associated with the direct gap (A-exciton, $E_{1s}^A$), and the exciton associated with the spin-orbit splitting of the direct gap valence band (B-exciton, $E_{1s}^B$) occurring at the K/K' points in the Brillouin zone, similar to the features reported in monolayer $MoS_2$ [9,10]. By measuring to an energy of 3.3 eV, we observe distinct spectral features we identify as the *2s* excited state, $E_{2s}^A$, of the A-exciton in both $WS_2$ and $WSe_2$. From these energy levels we derive a lower bound for the exciton binding energy as $E_{2s}^A - E_{1s}^A$ of 0.83 eV and 0.79 eV for $WS_2$ and $WSe_2$, respectively, and a lower bound for the corresponding band gaps $E_g \geq E_{2s}^A$ of 2.90 and 2.53 eV at 4 K. These values compare well with theoretical predictions (binding energies of 1.04 eV and 0.9 eV for $WS_2$ and $WSe_2$, respectively) [20]. Our experimental values are significantly larger than the 0.37 eV binding energy recently reported by He *et al* [23] for $WSe_2$, but are similar to the 0.7 eV binding energy reported by Ye et al [24] for $WS_2$. Both of these groups use two-photon photoluminescence spectroscopy measurements. A third group reported a binding energy of 0.32 eV in $WS_2$ [25]. A summary of the dispersion of theoretical and experimental results from these materials is presented in the *Supplementary Information*.

These remarkably high exciton binding energies imply that excitonic behavior dominates to room temperature and above, and we are indeed able to follow the evolution of these features to 300 K. Because the exciton binding energies are large, the true band gap is substantially higher than the dominant spectral feature commonly observed with photoluminescence and referred to as the "optical band gap" [9,10]. This information is critical for developing future technological applications, and enables further progress in both experimental and theoretical efforts to better understand these novel monolayer semiconductor materials.

## 2. Experimental Details

### 2.1 Sample preparation and characterization.

Samples are prepared by mechanically exfoliating flakes from bulk single crystals [26] and depositing them onto either Si/285nm $SiO_2$ or Si/90nm $SiO_2$ substrates. Candidate single layer regions are first identified with an optical microscope as shown in Fig. 1(a,b). The single layer nature of these regions is then confirmed based on the energy separation of the Raman features at room temperature. The Raman spectra showing separation of the $E_{2g}^1$ and $A_{1g}$ vibrational modes as measured with micro-Raman spectroscopy are shown in Fig. 1(c,d). For $WS_2$ the in-plane $E_{2g}^1$ and out-of-plane $A_{1g}$ Raman modes come closer together as the number of layers is decreased, just as in $MoS_2$. In the single layer limit, the energy difference is 60-61 $cm^{-1}$ while for the bulk it is 65 $cm^{-1}$ [11,27]. For $WSe_2$ the bulk in-plane and the out-of-plane modes are degenerate (E = 250 $cm^{-1}$), but this degeneracy is lifted for few and single layers. The resulting splitting is 11 $cm^{-1}$ for a single layer and 6 $cm^{-1}$ for the bilayer [27]. Fig. 1(c) shows a splitting of 60 $cm^{-1}$ for $WS_2$ and Fig. 1(d) shows a splitting of 12 $cm^{-1}$ for $WSe_2$, thus verifying the single layer nature of our flakes. Monolayer regions are typically 5–10 μm across.

### 2.2 Spectroscopy measurements.

To probe the electronic transitions in these monolayers, light from a broad-band radiation source is focused on the sample (the spot size of ~ 1 um diameter is depicted in Fig. 1(b)) and the reflected intensity is measured. This is commonly referred to as reflectivity. However, because the flakes are atomically thin and placed on $SiO_2/Si$, the reflected light spectra include the illumination source profile, the sample transmission and absorption, the substrate reflection and absorption, and resonant effects due to the thickness of the $SiO_2$. To distinguish the contribution from the monolayer TMD flake, we take the difference between the intensity measured from the flake, $I_{on}$, and from the substrate just off the flake, $I_{off}$, and normalize to the intensity from the substrate, $I^* = (I_{on} - I_{off}) / I_{off}$.

We used a micro-Reflectivity/PL setup (spatial resolution of 1 μm) with a 50x objective, appropriate filters and incorporating a continuous-flow He-cryostat to collect reflectivity and PL in a backscattering geometry. We performed baseline PL measurements at each temperature to verify the alignment of the sample with a continuous-wave 532 nm (2.33 eV) solid-state laser. For the absorption/reflectivity measurements, samples were excited with a W-halogen lamp fiber-coupled to the system. Emitted light was dispersed by a single-pass, 1/3 meter monochromator equipped with a multichannel charge coupled device (CCD) detector. The lamp output and spectral response of the system (including the CCD) enabled reliable reflectivity measurements to 3.3 eV. Reflectivity measured from different monolayer flakes on the same substrates all behave identically. Further details of the sample preparation and experimental spectra are available elsewhere [28].

## 3. Results and Discussion

### 3.1 Reflectivity spectra.

Spectra from $WS_2$ and $WSe_2$ at 4 K are shown in Fig. 2(a) and 2(b), respectively. We label features commonly identified as the direct-gap ground-state exciton occurring at the K/K' points of the band structure (A-exciton), and with the corresponding transition between the spin-orbit split-off valence band to the conduction band (B-exciton). Table I summarizes our $1s^A$ and $1s^B$ exciton energies. These values are similar to those reported [29]. The derivative-like behavior of the $WSe_2$ $1s^A$ as well as the ~500-meV-wide background peak near 2.2 eV are discussed in references 23, 30, and the *Supplementary Information*.

The measured energy difference between the $1s^A$ and $1s^B$ excitons is the spin-orbit splitting energy, $\Delta_{SO}$, with values of 391 meV and 412 meV for $WS_2$ and $WSe_2$, respectively. These compare well with theoretical values of 426 meV and 456 meV for $WS_2$ and $WSe_2$, respectively [20,31].

In addition to the A and B $1s$ features, we observe distinct, well-resolved features labeled Y at much higher energies (2.910 ± 0.002 eV and 2.533 ± 0.004 eV for $WS_2$ and $WSe_2$, respectively) with linewidths and intensities comparable to those of the A and B features. For $WS_2$, the intensity of Y is approximately one tenth that of A. In $WS_2$ we also observe a feature labeled Z at 3.1 eV which we tentatively associate with the band edge continuum.

To more readily distinguish the spectral features, we calculate the numerical derivative, $dI^*/dE$, to suppress the contribution from the smoothly varying background. Fig. 2(c,d) show the resultant differential spectra for $WS_2$ and $WSe_2$ at 4 K with the features labeled as described.

**3.2 Temperature dependence.**

Fig. 3 and Fig. 4 summarize the temperature dependence from 4-300 K over the full energy range of data collected. The spectra are offset for clarity, and fiducial marks at the positions of the $1s^A$ and $1s^B$ excitons and the Y features are a guide to show the shift of each with temperature. A magnification of the WS$_2$ spectra in the higher energy regime is shown in Fig. 3(b). The temperature dependence of the energy positions for the spectral features of WS$_2$ (WSe$_2$) are plotted in the *Supplementary Information*. All of the features shift monotonically to lower energy with increasing temperature, as expected for a semiconductor. More importantly, the temperature dependence is the same for all of the features, indicating a common origin.

**3.3 Identification of spectral features.**

As noted above, feature Y exhibits the same temperature dependence as the ground state exciton features A ($1s^A$) and B ($1s^B$) observed in WS$_2$ and WSe$_2$, and has a comparable linewidth as seen more clearly in Fig. 5. Gaussian fits yield linewidths of 25, 40 and 50 meV for the A, B and Y features in WS$_2$, and 40 and 45 meV for the B and Y features in WSe$_2$.

Several mechanisms could increase the absorption above the B exciton and produce a broad spectral feature (typically labeled "C") with stronger absorption than A or B. These mechanisms include band nesting [32] and nearly degenerate transitions near Γ which are broadened by electron-phonon interactions [33]. Although we see a modest increase in absorption above the B transition, we observe a clear, narrow-linewidth Y feature not expected from either of these effects. This Y feature is observed in both WS$_2$ and WSe$_2$, as shown in Figs. 2 and 3. We have observed this Y feature in several different monolayer samples of distinct origin, and find it to be independent of sample origin, preparation or substrate.

To further guide identification, we consider contributions from vertical band-to-band transitions expected to occur at critical points in the calculated band structures (*e.g.* near Γ or Q, see *Supplementary Information*) [20, 27, 31, 34]. The energies of these transitions relative to the $1s^A$ feature arising from the direct gap at the K/K' point are indicated by the shaded boxes along the horizontal axes of Fig. 3(a,b) and 4(a), with the box width determined by the range of calculated results. The calculated spin-orbit splitting energy $\Delta_{so}$ leading to the well established $1s^B$ exciton feature agrees very well with our low temperature data – the shaded box falls within the experimental linewidth of B. However, the energy of feature Y does not correspond to transitions at the Q and $\Gamma$ points for either $WS_2$ or $WSe_2$ (shaded boxes labeled $E_Q$ and $E_\Gamma$).

Since we are unable explain feature Y by band nesting [32] or other Brillouin-zone transitions, we thus consider its identification as the *2s* excited state of the A ($1s^A$) or B ($1s^B$) exciton. With reference to the $WS_2$ spectra, if we identify Y as $2s^B$ as 435 meV above B, we would also expect to see a $2s^A$ feature. In this scenario, such a $2s^A$ peak would appear at least 435 meV above A, occurring at 2.513 eV, just above the B-exciton peak. However, there is clearly no feature there. The clean lineshapes of the B feature in both the *I\** and derivative spectra, indicate that such a $2s^A$ peak is not hidden within the B linewidth.

He *et al* observe a weak-amplitude Rydberg series of excited states for the A exciton [23]. We also see these features but do not associate them with the sample response (see *Supplementary Information* S2). Based on all of these factors, we conclude that feature Y is the first excited state $2s^A$ of the A exciton.

With this identification, these data establish a lower bound for the exciton binding energy free of any model, given by $E_{2s}^A - E_{1s}^A$ of 0.83 eV and 0.79 eV for $WS_2$ and $WSe_2$, respectively. The energy $E_{2s}^A$ also provides a lower bound for the corresponding low temperature band gaps $E_g \sim E_{2s}^A$ of 2.91 and 2.53 eV. These values compare well with those predicted by first principles theory based on the *GW* approximation in conjunction with the Bethe-Salpeter equation (binding energies of 1.04 eV and 0.9 eV, and band gaps of 2.88 and 2.42 eV for $WS_2$ and WSe$_2$, respectively) [20]. A compilation of experimental and theoretical binding energy values for WS$_2$ and WSe$_2$ from various sources shown in *Supplementary Information* Fig. S4 reveals a wide variation in results, suggesting that further work is necessary to reliably identify spectral features and more accurately determine the band gaps and exciton binding energies.

### 3.4. Discussion.

It is instructive to discuss these results in the context of the well-known 2D hydrogen model for excitons [35,36]. Such a model has been used to gain insight into the monolayer TMDs, although it has been noted that the reduced dielectric screening in these monolayers may limit the applicability of this model [19,20]. Energy eigenvalues are given by:

$$E_n = -\frac{Ry}{(n-1/2)^2}, \quad (1)$$

where $n = 1, 2, 3, \ldots$ is the principal quantum number and $Ry = \mu m_o e^4/2\hbar^2(4\pi\varepsilon_o)^2\kappa^2$ is the effective Rydberg constant. Here $\mu = m_e m_h/(m_e + m_h)$ is the reduced mass of the electron ($m_e$) and hole ($m_h$) and $\kappa$ is the effective dielectric constant. The other variables are defined in the customary way. The optical absorption is proportional to $1/(n-1/2)^3$ [35,36], which predicts the relative intensity of the *2s* to the *1s* exciton absorption as 1/27. Our data, *I\**, is proportional to the absorption of the system [37]. Fig. 5 shows the normalized absorption of WS$_2$ at (a) 300 K and (b) 4 K obtained by removing a smoothly varying background and normalizing the spectrum to the A-exciton peak intensity. The ratio of the *2s$^A$* to *1s$^A$* peak intensity we measure is approximately 1/8 at 4 K and 1/30 at 300 K.

The exciton binding energy and band gap can also be predicted within this model from the experimental values of the exciton energies $E_{1s}^A$ and $E_{2s}^A$. The binding energy is given by $BE_{1s}^A = 9/8\,(E_{1s}^A - E_{2s}^A)$, and is 0.929 eV and 0.887 eV in WS$_2$ and WSe$_2$ respectively, in good agreement with our model-free lower bounds, as well as with recent theory values of 1.04 and 0.9 derived from a Bethe-Salpeter equation approach [20]. These binding energies lead to band edge energies $E_{gap} = E_{1s}^A + BE_{1s}^A$ of 3.01 eV and 2.63 eV for $WS_2$ and $WSe_2$. Given the experimental values $E_{1s}^A$ and $E_{2s}^A$, the other excited states and band edge continuum predicted by the 2D hydrogen model are shown superposed on the normalized absorption of $WS_2$ at 4 K in Fig. 5(b). A very reasonable correspondence is observed.

## 4. Conclusion

In conclusion, by examining optical reflectivity/absorption data over a wide range of energies and temperatures, we identify the *2s* excited state of the A-exciton, enabling experimental determination of exciton binding energies for single layers of $WS_2$ and $WSe_2$. Because the binding energies are large, the true band gap is substantially higher (~ 1 eV) than the dominant excitonic spectral feature commonly observed with photoluminescence, and the exciton can be directly created using photoexcitation energies well below the true band gap. This has been an implicit, and to date largely unappreciated assumption of much experimental work. In contrast with most common semiconductors, excitons in these novel monolayer materials exist even at room temperature, and an excitonic band structure and framework should be used as a starting point to further understand their basic properties [38]. More advanced theoretical treatments [20,21] may provide more accurate estimates of the exciton binding energies, band gaps, and corresponding properties when coupled with our experimental measurements of the *2s* exciton states. It is clear from the broad dispersion of results in the community that further work is necessary to fully understand these novel 2D materials.

**Acknowledgments**

We thank Jim Culbertson for assistance with Raman measurements. GK gratefully acknowledges the hospitality and support of the Naval Research Laboratory where the experiments were performed. This work was supported by ONR, core programs at NRL and the NRL Nanoscience Institute.

**Figure captions**

**Figure 1.** Optical microscope images of (a) WS$_2$ and (b) WSe$_2$. Raman spectrum of the monolayer regions taken at 300 K with an excitation energy of 488 nm are shown for (c) WS$_2$ and (d) WSe$_2$.

**Figure 2.** Relative change in intensity, $I^*$, from (a) WS$_2$ and (b) WSe$_2$ at 4 K. Panels (c) and (d) show the numerical derivatives $dI^*/dE$ of these spectra. The region in the rectangle in (c) is discussed in the *Supplementary Information*.

**Figure 3.** Differential intensity ratio spectra for WS$_2$ (a) from 4 K to 300 K over the full energy range of the data. (b) a magnification for the higher energy feature and band edge. The shaded boxes indicate energy ranges over which vertical transitions are predicted to occur.

**Figure 4.** Differential intensity ratio spectra for WSe$_2$ from 4 K to 300 K over the full energy range of the data. The shaded boxes indicate energy ranges over which vertical transitions are predicted to occur.

**Figure 5.** Normalized absorption spectra for WS$_2$ (a) at 300 K and (b) 4 K. The main features include the spin-orbit energy, $\Delta_{SO}$, as well as other features described in the text. Panel (b) also includes a comparison of WS$_2$ spectral features with a 2D hydrogen model.

|  | Peak Energies (eV) Experimental (4K) | | | $\Delta_{SO}$ (eV) | Binding Energy (eV) | |
|  | $E_A$ | $E_B$ | $E_Y$ | $E_B - E_A$ | Lower Bound $E_Y - E_A$ | Theory ref. [20] |
|---|---|---|---|---|---|---|
| $WS_2$ | 2.078 | 2.475 | 2.910 | 0.391 | 0.83 | 1.04 |
| $WSe_2$ | 1.740 | 2.162 | 2.533 | 0.412 | 0.79 | 0.90 |

**Table 1.** Experimentally determined peak positions and binding energies.

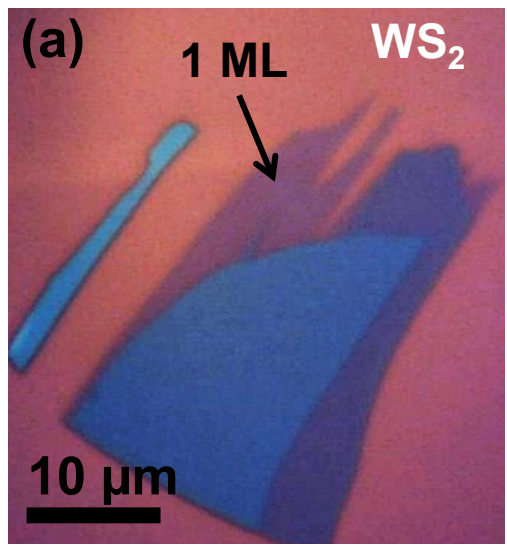
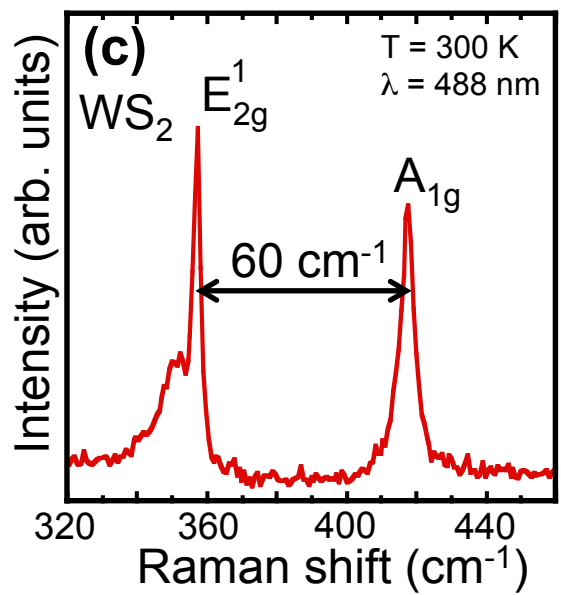
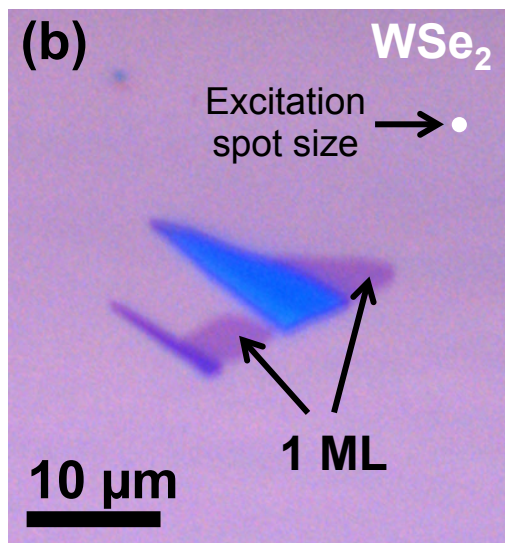
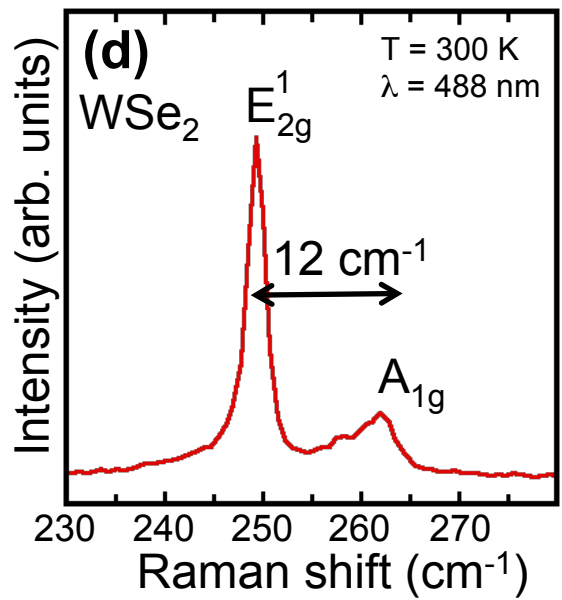

Figure 1

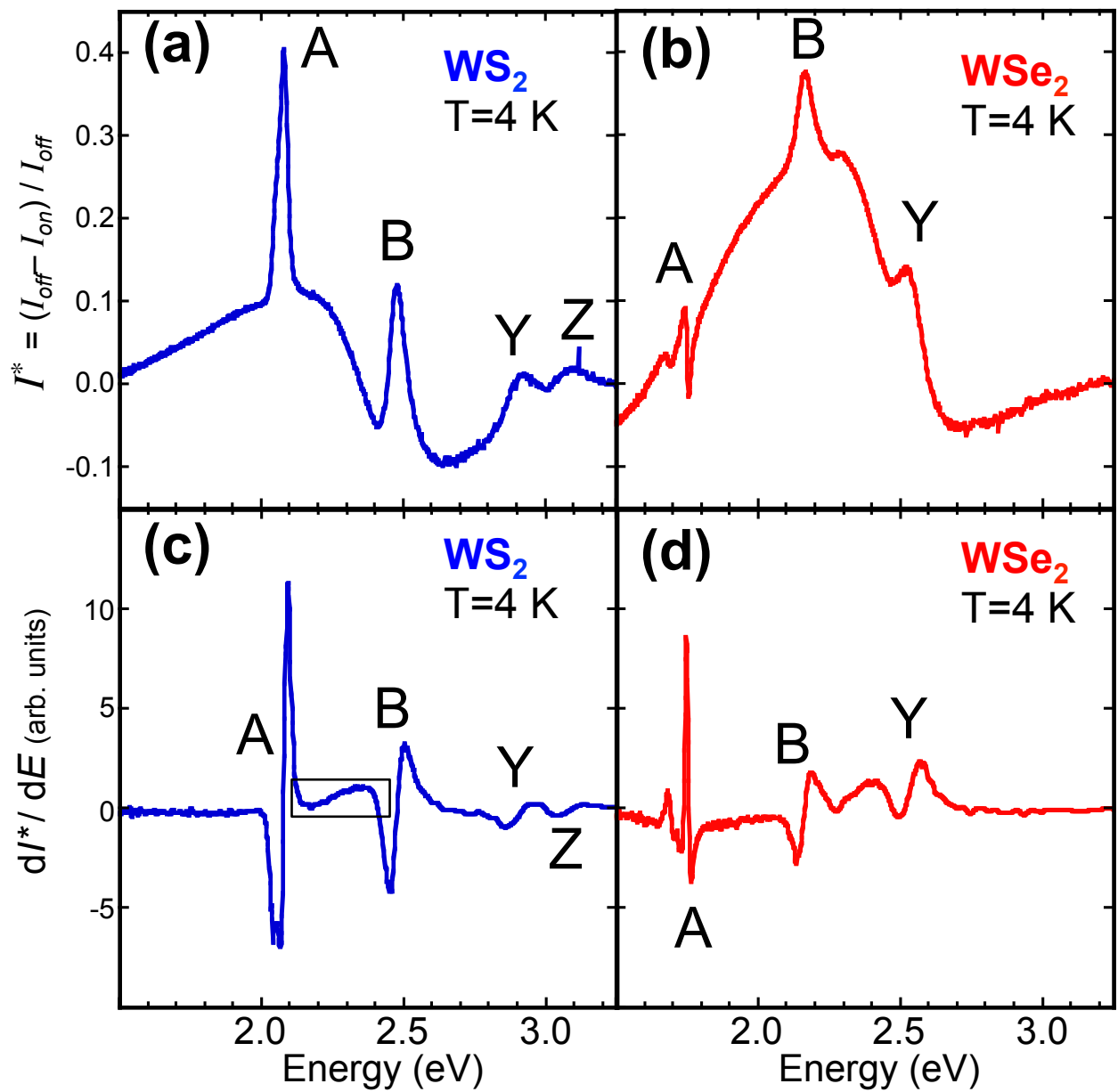

Figure 2

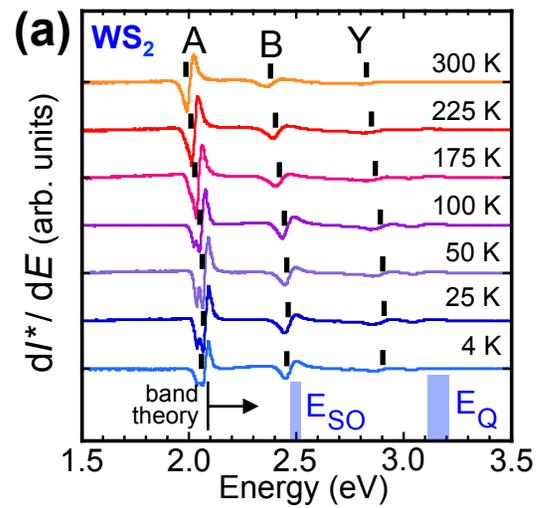
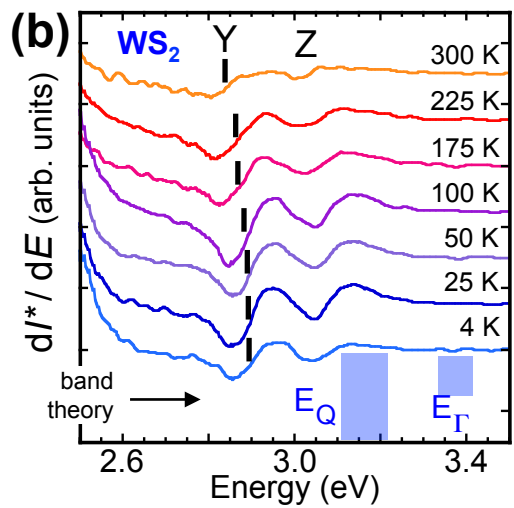

Figure 3

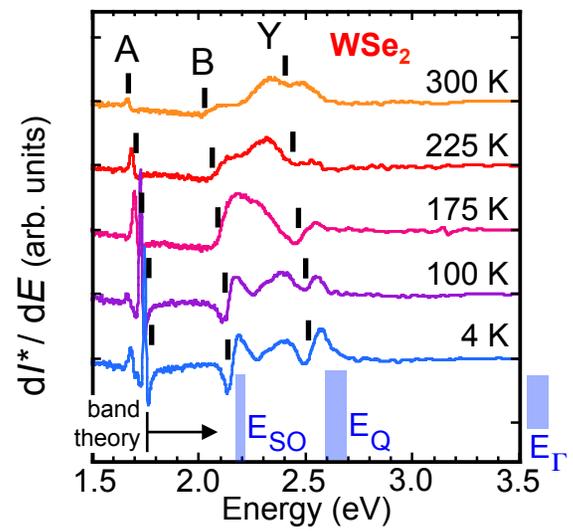

Figure 4

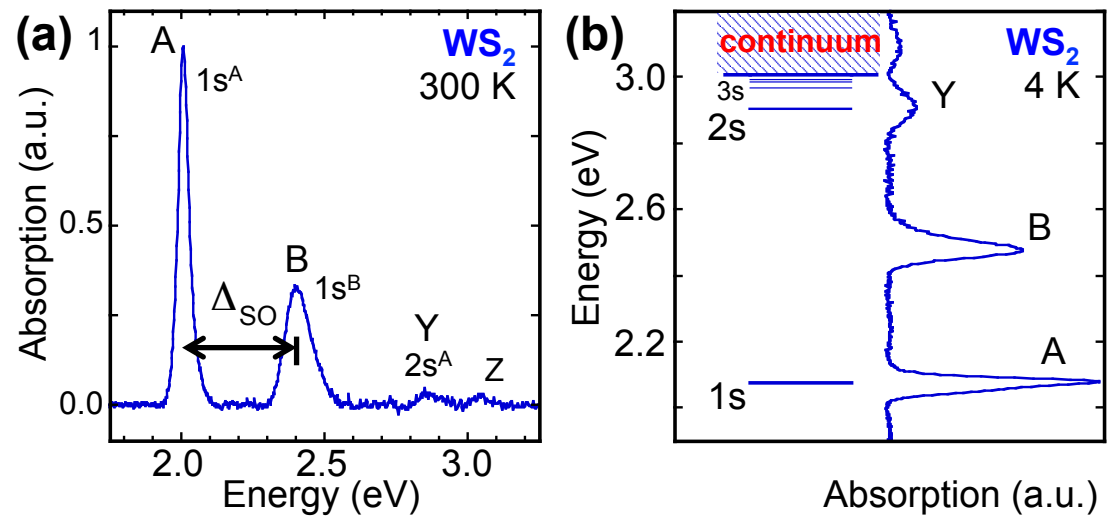

Figure 5

# Supplemental Information

# Measurement of high exciton binding energy in monolayer transition-metal dichalcogenides $WS_2$ and $WSe_2$

A. T. Hanbicki[a], M. Currie[a], G. Kioseoglou[b], A. L. Friedman[a], and B. T. Jonker[a*]

[a]*Naval Research Laboratory*, 4555 Overlook Ave. SW, Washington, DC 20375
[b]*Dept. of Materials Science and Tech., University of Crete*, Heraklion Crete, 71003, Greece

**S1. Band structure and vertical transitions at critical points.**

While we observe significant absorption at the K-point in the Brillioun zone, non-zero absorption may also occur at points in the Brillioun zone with a vertical transition and a significant density of states, e.g. where the band dispersion is low, as at a critical point. We identify such potential transitions from published band structures [1,2,3,4]. An example, shown in Fig. S1(a) is the single particle band structure for $WS_2$ from Zhu, et al. [1]. From this figure it is clear that the lowest energy gap is a direct transition at the K-point. There are also vertical transitions indicated by arrows in the figure between regions of the band structure with low or zero dispersion in both the conduction and valence bands at the Γ-point, and roughly halfway between Γ-K near the Q-point. A definition of the symmetry points in the Brillioun zone is shown in Fig. S1(b).

Although the absolute value of the direct gap at the K-point varies widely in the literature [1,2,3,4] due to the well known tendency for density functional computations to underestimate the band gap, the band energies relative to the K-point are very consistent among the published band structures. Therefore, we normalize to the direct band gap energy at the K-point by calculating the energy difference between a vertical transition at a given critical point and that at the K-point. Applying this procedure to the published band structures [1,2,3,4], we can identify a range of energies relative to the position of our experimental A-exciton ($E_{1s}^A$) peak of Figure 2 where we may expect a peak to occur corresponding to absorption at the Γ- or Q-points. These ranges are indicated by the shaded boxes along the bottom, horizontal axes of Fig. 3a, Fig. 3b, and Fig. 4. The spin-orbit energy predicted in these calculations is also indicated in these figures as a consistency check and agree quite well with the observed experimental spectra.

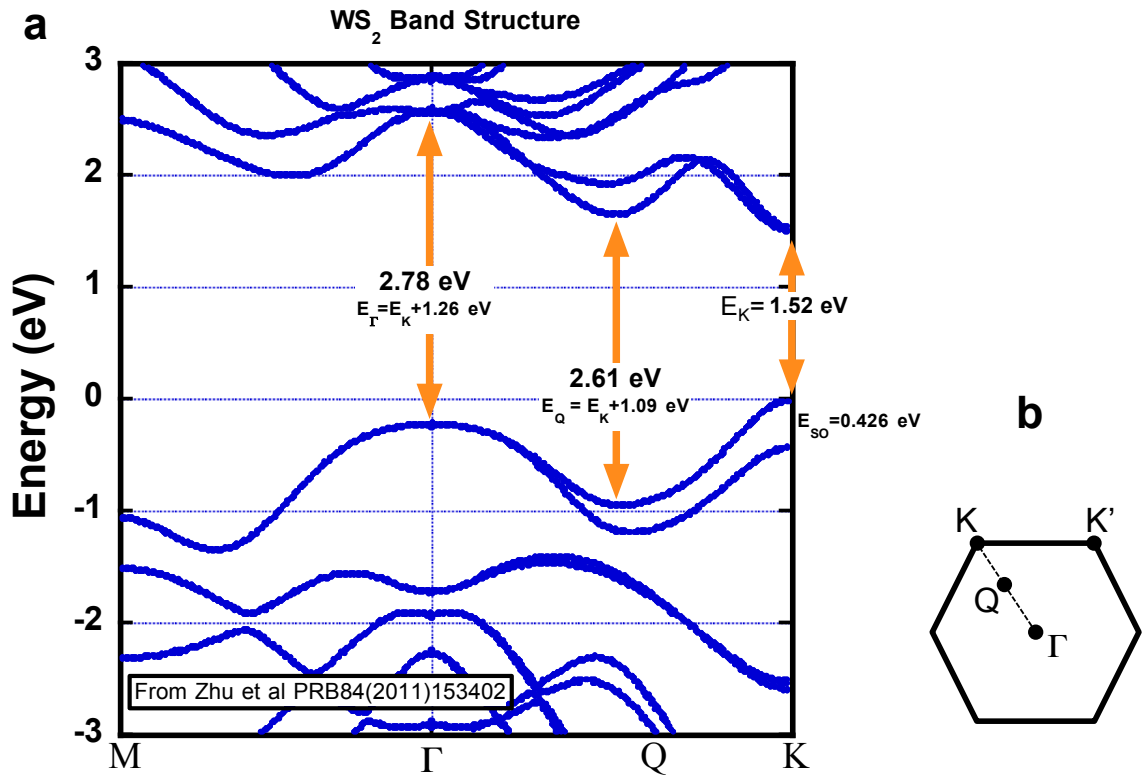

**Figure S1.** (a) Band structure from reference [1] and (b) location of various high symmetry points in the Brillioun zone. The parameters derived from this figured and used to compare with our experimental data are the differences between the absolute gap, $E_K$, and the energy at various high symmetry points.

## S2. Low intensity oscillations in the absorption spectra

We note that He *et al.* observe a series of small features in the magnified second derivative spectra between the A and B peaks which they identify as a Rydberg series of excited states for the A exciton [5], in analogy with such series observed in bulk crystals due to delocalized excitons [6]. They note that the level spacing is more even than expected from a 2D hydrogenic model, but deduce a binding energy of 0.37 eV. We also see such features between the A and B peaks in our magnified second derivative data, and find that they are also present in control spectra recorded from a spot not on the WS$_2$ sample, *i.e.* from the SiO$_2$/Si substrate. To ensure we are not missing features hidden in the noise of our data, we have taken care to fully scrutinize our spectra. Figure S2 presents a magnification of the data shown in Fig. 2(c) of the manuscript for monolayer WS$_2$ over the energy range between the A and B exciton peaks (2.1 to 2.4 eV, blue trace). The area indicated by a box in Fig. 2(c) is a shown here, magnified by a factor of roughly 8.5x. When significantly magnified in this way, oscillations become apparent in the data. However, similar oscillations are observed in a spectrum acquired from a spot completely off of the WS$_2$ sample, i.e. from the SiO$_2$ / Si substrate (Fig. S2, red trace). Therefore, we do not associate them with the electronic structure of the WS$_2$ monolayer. The oscillations appear more evenly spaced than expected from a Rydberg series, with an energy spacing on the order of ~25-50 meV. Based on the diffraction relation, $d = \frac{1}{2(\Delta E)n}$, this energy spacing corresponds to a characteristic length scale of ~5-15 $\mu$m assuming an index of refraction, $n$, near 2 ($n_{Si} = 3.5$, $n_{SiO2}=1.5$). There are many components in the experimental setup that can contribute to such a background. These include various filters and lenses, as well as a possible etalon effect in the CCD detector used in the spectrometer.

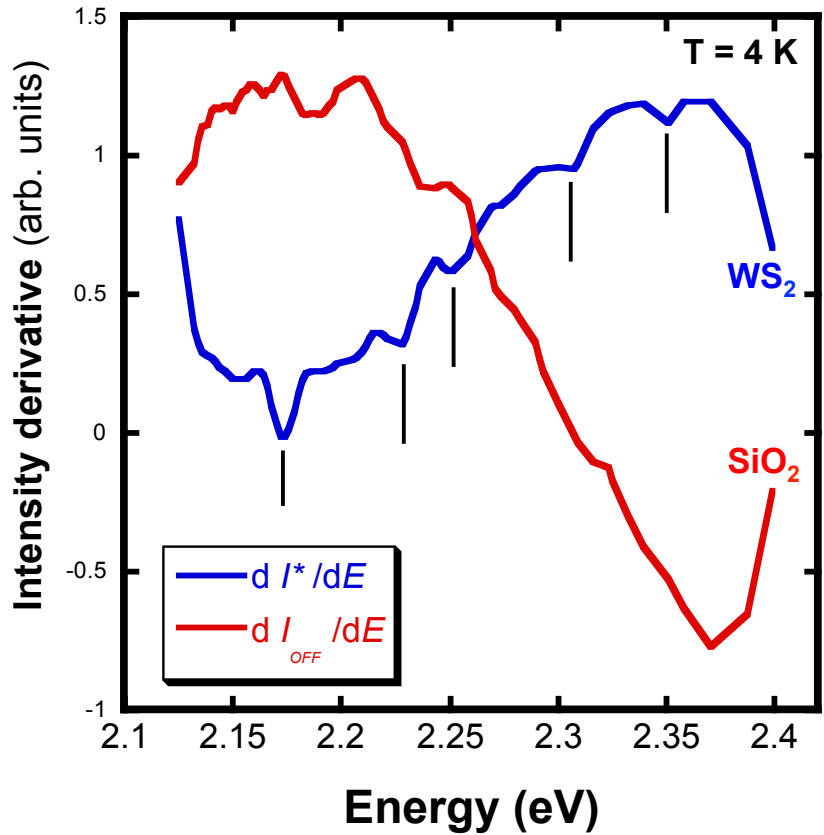

**Figure S2.** Magnification of the derivative of measured intensity. The blue trace is a ~8.5x magnification of $I^*$, the data presented in Fig. 2(c), between the A- and B-exciton. The red trace is numerical derivative of $I_{off}$, the intensity from the substrate in the same energy window. Fiducial marks indicate the positions of several of the more pronounced oscillations, although others may be identified.

## S3. Temperature dependence

The temperature dependence of the energy position of the A ($1s^A$), B ($1s^B$) and Y features of Figure 2, 3, and 4 can be described by the relation $E(T) = E(0) - S\langle\hbar\omega\rangle[\coth(\langle\hbar\omega\rangle/2kT) - 1]$ where E(0) is the energy position of a given feature at zero temperature, S is a dimensionless coupling constant, and $\langle\hbar\omega\rangle$ is an average phonon energy [7]. From the fits presented in Fig. S3(a) and S3(b), we extract an average phonon energy of $\langle\hbar\omega\rangle$ = 14 meV (15 meV), and S = 1.84(2.11), 2.02(2.71), and 1.37(2.34) for the A, B, and Y features, respectively for $WS_2$ ($WSe_2$). The values for $\langle\hbar\omega\rangle$ and S for the A and B features are similar to those reported for $MoS_2$ [8].

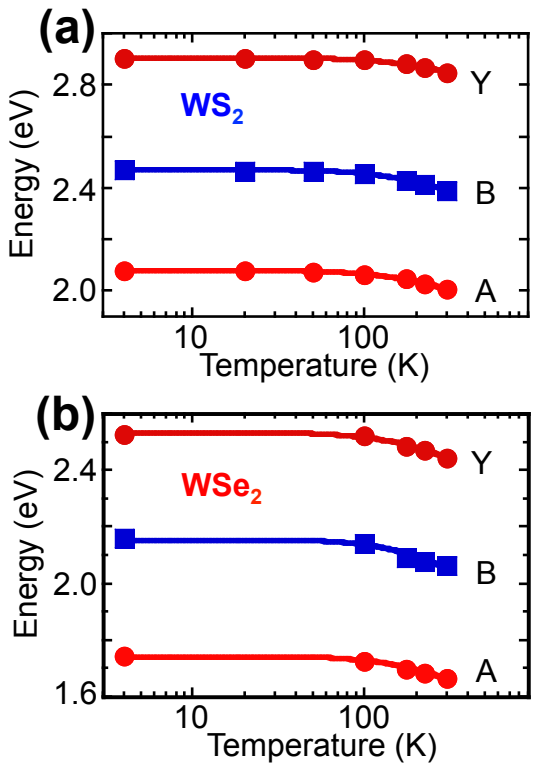

**Figure S3.** Summary of the peak positions versus temperature for (a) $WS_2$ and (b) $WSe_2$. Solid lines are fits from the equation discussed in the text.

**S4. Substrate effects on the spectra**

The thin-film TMD materials are placed on a silicon substrate with a thin (80 – 250 nm) silicon dioxide layer on top. This creates a weak optical cavity due largely to the reflectivity of the Si/SiO$_2$ interface. To understand any resonant effects, such as those introduced by an optical cavity, we model several scenarios with the physical characteristics of our systems to give us insight on the cavity properties. For simplicity we do not include the monolayer in these models. Figure S4(a) presents some simulations of a Fabry-Perot resonator based on these materials using the wavelength-dependent reflectivity for silicon and dispersion-free silicon dioxide. From this series of simulations, we observe that thinner SiO$_2$ layers have a less complex reflectivity spectrum in the energy range we are probing. An 80-nm-thick SiO$_2$, for instance, is almost a linear function of photon energy in the range of 1-3 eV. On the other hand, a 290-nm-thick SiO$_2$ layer has an oscillatory reflectivity that is peaked near 2 eV, in the center of our region of interest. Such cavity effects in conjunction with the absorption of our flakes likely contribute to the smoothly varying background seen in the data presented in Fig. 2(a) and 2(b).

In addition to possible contributions to the overall background, another noteworthy effect occurs when loss is included in the simulation. In our model we added two Gaussian absorption peaks, as shown in Fig. S4(b). The cavity resonance properties dictates that absorption can manifest itself as either a peak, a valley, or a combination that is reminiscent of the derivative-like lineshape we observe for our WSe$_2$ data in Fig. 2(b). It is interesting to note that near an inflection point in the resonator's response the absorption peak nearly disappears. Thus, careful attention to the design of cavity properties are needed to ensure that data can be extracted in a straightforward way.

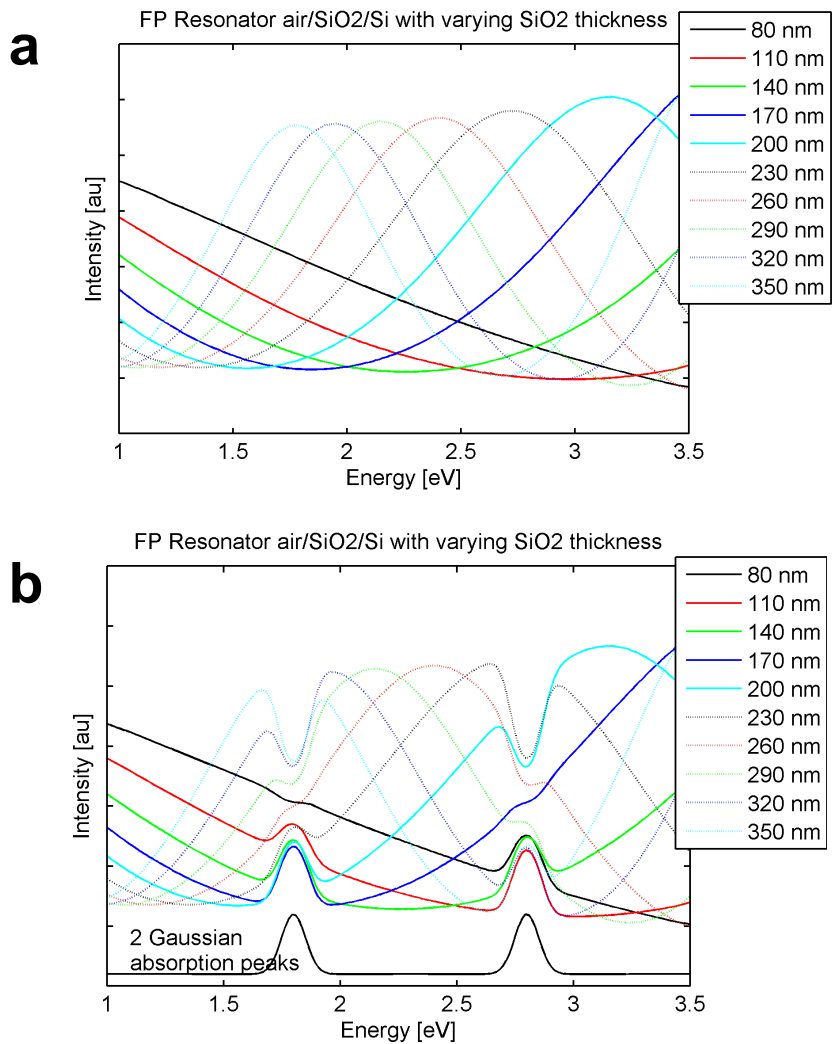

**Figure S4.** (a) Cavity simulation of an air/SiO$_2$/Si system using a Fabry-Perot resonator model. These simulations are presented for a series of SiO$_2$ thicknesses. (b) Two Gaussian absorption peaks are included in the cavity resonance simulations to illuminate the effect of cavity resonance on absorption peak lineshapes.

## S5. Comments on the 2D hydrogen model

The exciton wave functions in these materials have large spatial extent, and are therefore Wannier-like in character [9,10,11] consistent with the 2D hydrogen model despite the reduced dielectric screening. The parameters derived from our experimental data, coupled with the effective exciton mass of reference 9, yield Bohr radii of 8.6 Å and 8.2 Å, which are larger than the in-plane lattice parameters of 3.2 Å and 3.3 Å for WS$_2$ and WSe$_2$, respectively. We note that the radius of the *2s* excited state is even larger than that of the *1s* ground state. Further evidence of the extended nature of the exciton wavefunction can be found in recent papers on the TMD alloy $MoS_{2(1-x)}Se_{2x}$ [12,13]. The authors measure a continuous shift of the exciton emission energy as a function of alloy concentration. This demonstrates that the exciton's spatial extent is on the order of a few unit cells, because otherwise one would expect the measured spectra to consist of a superposition of the individual exciton peaks characteristic of $MoS_2$ and $MoSe_2$. These length scales imply Wannier-like excitons, providing some modest justification for the hydrogen model.

**S6. Summary of binding energies found in the literature**

Figure S5 is a summary of experimental reports and theoretical predictions of the binding energy for $WS_2$ and $WSe_2$ [2,4,14,15,16,17,18,19,20] . The binding energy values obtained in this work are indicated by the open diamond symbols. They are model independent lower bounds which depend upon identification of the exciton ground (1s) and first excited state (2s) in the absorption spectra, with a lower bound for the binding energy given by the energy difference between these features. The figure identifies whether the value reported is from theory only (squares), or from experiment coupled with theory (closed circles). Note that there is one data point from experiment only, independent of a model (open circle). There is a wide variation in the reported values for both experiment and theory.

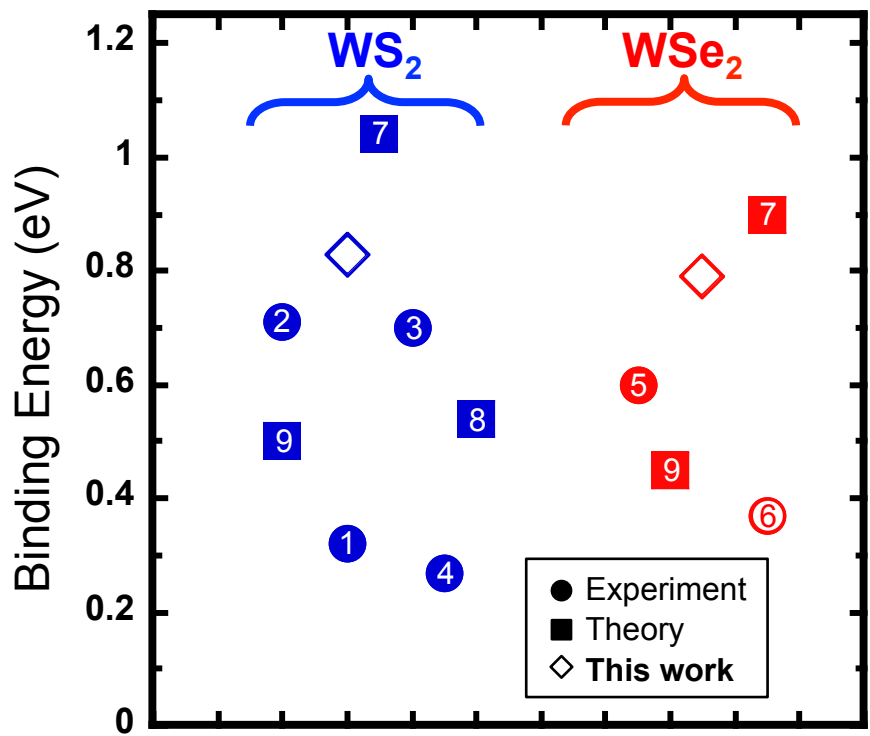

**Figure S5.** Summary of exciton binding energies reported for monolayer $WS_2$ (blue) and $WSe_2$ (red). Values obtained in this work are indicated by the open diamond symbols. Squares symbols represent theoretical predictions, closed circles are results of experiment coupled with theory, and the open circle is from a model independent experimental work. The numbers on the points correspond to the following references:
1. A. Chernikov *et al.*, *PRL* **113**, 076802 (2014)
2. B. Zhu *et al.*, arXiv:1403.5108
3. Z. Ye *et al.*, *Nature* **513,** 214 (2014)
4. T. Stroucken and S.W. Koch, arXiv:1404.4238
5. G. Wang *et al.*, arXiv:1404.0056
6. K. He *et al.*, *PRL* **113**, 026803 (2014)
7. A. Ramasubramanium, *PRB* **86**, 115409 (2012)
8. H. Shi *et al.*, *PRB* **87**, 155304 (2013)
9. T.C. Berkelbach *et al.*, *PRB* **88**, 045318 (2013)